\documentclass{ifacconf}

\usepackage{graphicx}      % include this line if your document contains figures
\usepackage{natbib}        % required for bibliography
\usepackage{amsmath}
\usepackage{amssymb}
\usepackage{mathtools}
\usepackage{mleftright}
\usepackage[utf8]{inputenc}
\usepackage{prettyref}
\usepackage{xcolor}
\usepackage[permil]{overpic}
\usepackage{booktabs}
\usepackage{multirow}
\usepackage{ifthen}
\usepackage{xspace}
\usepackage{siunitx}
\usepackage{nicefrac}
\usepackage{subfigure}
\usepackage{tikz}
\usepackage{pgfplots}
\usetikzlibrary{external}
\usetikzlibrary{pgfplots.groupplots}
\allowdisplaybreaks

\DeclareMathOperator\erf{erf}
\newcommand{\dd}{\mathrm{d}}
\newcommand{\Diff}[2]{\frac{\dd#1}{\dd#2}}

\newcommand{\DDiff}[2]{\frac{\dd^2#1}{\dd#2^2}}
\newcommand{\DDiffT}[1]{\DDiff{#1}{t}}
\newcommand{\PartDiff}[2]{\frac{\partial #1}{\partial #2}}

\newcommand{\real}[1]{\textrm{Re}\left\{#1\right\}}

\DeclareRobustCommand{\vec}[1]{ 				
	\ifthenelse{\equal{#1}{\omega} \OR \equal{#1}{\varphi} \OR \equal{#1}{\alpha} \OR \equal{#1}{\beta} \OR \equal{#1}{\chi} \OR \equal{#1}{\delta} \OR \equal{#1}{\varepsilon} \OR \equal{#1}{\phi} \OR \equal{#1}{\epsilon} \OR \equal{#1}{\gamma} \OR \equal{#1}{\eta} \OR \equal{#1}{\iota} \OR \equal{#1}{\kappa} \OR \equal{#1}{\lambda} \OR \equal{#1}{\mu} \OR \equal{#1}{\nu} \OR \equal{#1}{\pi} \OR \equal{#1}{\theta} \OR \equal{#1}{\vartheta} \OR \equal{#1}{\rho} \OR \equal{#1}{\sigma} \OR \equal{#1}{\varsigma} \OR \equal{#1}{\tau} \OR \equal{#1}{\upsilon} \OR \equal{#1}{\xi} \OR \equal{#1}{\psi} \OR \equal{#1}{\zeta} \OR \equal{#1}{\mathcal{A}}  \OR \equal{#1}{\mathcal{B}} \OR \equal{#1}{\mathcal{C}} \OR \equal{#1}{\mathcal{D}} \OR \equal{#1}{\mathcal{E}} \OR \equal{#1}{\mathcal{F}} \OR \equal{#1}{\mathcal{G}} \OR \equal{#1}{\mathcal{H}} \OR \equal{#1}{\mathcal{I}} \OR \equal{#1}{\mathcal{J}} \OR \equal{#1}{\mathcal{K}} \OR \equal{#1}{\mathcal{L}} \OR \equal{#1}{\mathcal{M}} \OR \equal{#1}{\mathcal{N}} \OR \equal{#1}{\mathcal{O}}
 \OR \equal{#1}{\mathcal{P}} \OR \equal{#1}{\mathcal{Q}} \OR \equal{#1}{\mathcal{R}} \OR \equal{#1}{\mathcal{S}} \OR \equal{#1}{\mathcal{T}} \OR \equal{#1}{\mathcal{U}} \OR \equal{#1}{\mathcal{V}} \OR \equal{#1}{\mathcal{W}} \OR \equal{#1}{\mathcal{X}} \OR \equal{#1}{\mathcal{Y}} \OR \equal{#1}{\mathcal{Z}}}{
		\boldsymbol{#1}
	}{
		\mathbf{#1}
	}
}

\newrefformat{sec}{Section~\ref{#1}}
\newrefformat{fig}{Fig.~\ref{#1}}
\newrefformat{tab}{Tab.~\ref{#1}}
\newrefformat{the}{Theorem~\ref{#1}}
\newrefformat{lem}{Lemma~\ref{#1}}
\newrefformat{rem}{Remark~\ref{#1}}
\newrefformat{app}{Appendix~\ref{#1}}
\newrefformat{ass}{Assumption~\ref{#1}}
\newrefformat{pro}{Proposition~\ref{#1}}

\definecolor{acin_red}{RGB}{186, 18, 43}
\definecolor{acin_gray}{RGB}{176, 176, 176}
\definecolor{acin_yellow}{RGB}{252, 204, 71}
\definecolor{acin_green}{RGB}{0, 190, 65}
\definecolor{TU_blue}{RGB}{0, 102, 153}
\definecolor{test2}{rgb}{0,0.47,0.85}
\definecolor{TU_gray}{RGB}{102, 102, 102}
\providecolor{test}{rgb}{0.85, 0, 0.22}

\begin{document}
\begin{frontmatter}

\title{Optimal control of quasi-1D Bose gases in optical box potentials} 

\author[ACIN]{A. Deutschmann-Olek}
\author[ACIN]{K. Schrom}
\author[ACIN]{N. Würkner}
\author[ATI]{J. Schmiedmayer}
\author[ATI]{S. Erne}
\author[ACIN,AIT]{A. Kugi} 

\address[ACIN]{Complex Dynamical Systems Group, Automation and Control Institute, TU Wien, Vienna, Austria}
\address[ATI]{Vienna Center for Quantum Science and Technology, Atominstitut, TU Wien, Vienna, Austria}
\address[AIT]{Center for Vision, Automation \& Control, Austrian Institute of Technology, Vienna, Austria}

\begin{abstract}                % Abstract of not more than 250 words.
	In this paper, we investigate the manipulation of quasi-1D Bose gases that are trapped in a highly elongated potential by optimal control methods. The effective mean-field dynamics of the gas can be described by a one-dimensional non-polynomial Schrödinger equation. We extend the indirect optimal control method for the Gross-Pitaevskii equation by \cite{winckel_computational_2008} to obtain necessary optimality conditions for state and energy cost functionals. This approach is then applied to optimally compress a quasi-1D Bose gase in an (optical) box potential, i.e., to find a so-called short-cut to adiabaticity, by solving the optimality conditions numerically. The behavior of the proposed method is finally analyzed and compared to simple direct optimization strategies using reduced basis functions. Simulations results demonstrate the feasibility of the proposed approach.
\end{abstract}


\end{frontmatter}
%===============================================================================
%==  Content  =================================================================
%===============================================================================

% include sections
\section{Introduction and Motivation}
\label{sec:introduction}

Recent progress in quantum experiments vastly improved the capabilities of controlling and manipulating individual quantum systems and laid the foundation for future technologies that try to harness genuine quantum effects to fulfill desired tasks with superior performance. 
Observing and exploiting quantum effects experimentally is a notoriously hard problem due to the fragility of quantum states in their natural surrounding.
Suppressing disturbances and other detrimental effects on the system behavior is a classical task in control engineering.
As a consequence, methods from automatic control and control engineering have proven highly valuable to achieve such goals, see, e.g., \cite{grond_optimal_2009,dotsenko_quantum_2009,omran_generation_2019,magrini_real-time_2021}, and are expected to become vital building blocks for quantum technologies.
%A active research area are so-called quantum simulators \cite{georgescu_quantum_2014} aim to simulate quantum systems that are infeasible to calculate on conventional computers.
% Current research efforts towards quantum technologies increasingly try to utilize genuine quantum effects such as entanglement, squeezing and non-classical states to fulfil desired tasks with superior performance. For example, quantum sensing and metrology \cite{giovannetti_advances_2011,degen_quantum_2017} tries to improve classical measurement tools while quantum simulators \cite{georgescu_quantum_2014} aim to simulate quantum systems that are infeasible to calculate on conventional computers. The latter is especially relevant for quantum many-body dynamics that are at the core of many of the most intriguing problems in physics from the early universe to quantum materials. The application of methods from control 

A particularly versatile model platform to explore and utilize quantum many-body systems are trapped clouds of ultra-cold atoms that form Bose gases. The reliable manipulation of atomic clouds on Atom chips suggested the use of optimal control methods to optimize the handling of trapped Bose-Einstein condensate in one and three dimensions, see  \cite{grond_optimal_2009, mennemann_optimal_2015}. 
Recent interest in quantum-thermodynamic questions and the lack of suitable experimental platforms stimulated the investigation of thermal machines operating on quantum fields. 
Thereby, an ultra-cold atomic cloud in a highly elongated trap is used as a working fluid. This quasi-1D Bose gas is then manipulated in longitudinal direction, i.e., the weakly confined direction, to implement desired thermal operations. 
Since such operations introduce unwanted heat to the gas in general, it seems natural to use optimal control methods to mitigate such heating effects.
The finite temperature case of the quantum many-body system would require a description of the quantum field's stochastic fluctuations, which is mathematically quite involved. 
As a first step, this contribution therefore seeks to find solutions in terms of the system's mean field, which can be accurately described by the Gross-Pitaevskii equation (GPE) in three dimensions. However, finding optimal solutions of 3D problem as presented by \cite{mennemann_optimal_2015} is computationally very expensive.

The highly elongated setting exhibits a coupling of the transversal and longitudinal dynamics since atomic interactions lead to a transversal broadening of the gas. Nevertheless, one can effectively describe the resulting longitudinal mean-field dynamics of the Bose gas by a one-dimensional non-polynomial Schrödinger equation (NPSE), e.g., as derived in \cite{salasnich_effective_2002}. Since this nonlinearly affects the local propagation speed of excitations, it needs to be taken into account for optimal control solutions.
While methods to create arbitrary (time-varying) optical 1D potentials have been presented recently by \cite{deutschmann_iterative_2022} with first experimental results by \cite{calzavara_optimizing_2022}, simple optical box potentials are feasible with state-of-the-art algorithms presented in \cite{tajik_designing_2019}.

This paper investigates optimal control solutions for the longitudinal dynamics of a quasi-1D Bose gas described by the non-polynomial Schrödinger equation, which is briefly introduced in \prettyref{sec:npse}. After introducing the optimal control problem, \prettyref{sec:optimal_control} extends existing approaches for the 1D GPE given in \cite{winckel_computational_2008} to the required npSE. \prettyref{sec:compression} applies these solutions to optimally compress a Bose gas in a box potential and compares the resulting solution to simple basis-function approaches. Resulting conclusions and a short outlook are finally given in \prettyref{sec:outlook}.

\section{Mean-field dynamics of quasi-1D Bose gases}
\label{sec:npse}

\begin{figure}[htbp]
	\begin{center}
		\includegraphics[width=\columnwidth]{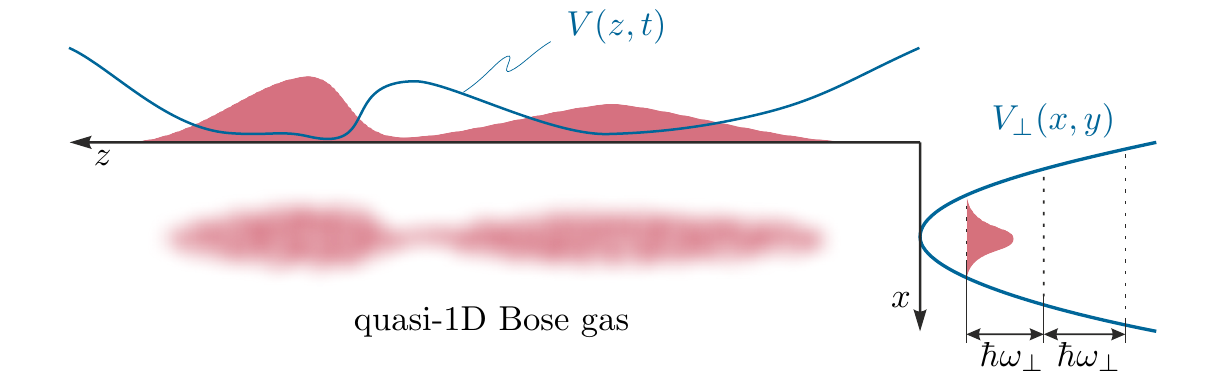}
		\caption{Density $\rho$ of an elongated atomic cloud with tight transversal confinement due to the static trapping potential $V_{\perp}(x,y)$ of the atom chip. Using spatially shaped attractive or repulsive light fields to create an additional optical dipole potential, one can generate complex time-varying longitudinal potentials $V(z,t)$.}
		\label{fig:atomchip}
	\end{center}
\end{figure}

We consider a cloud of $N$ ultra-cold atoms trapped in a highly elongated trap with a tight transversal confinement due to the static parabolic potential $V_{\perp}(x,y) = \frac{\omega_\perp}{2}  (x^2 + y^2)$, with the transversal angular frequency $\omega_{\perp}$.
Along the weakly confined longitudinal direction $z \in \mathcal{D} = [-L/2,L/2]$, a tunable potential $V(z,t)$ is created using spatially shaped light fields. 
We assume that the condensate remains in a Gaussian state close to the ground state in transversal direction, as illustrated in \prettyref{fig:atomchip}, which is the case for sufficiently low temperatures $T$, i.e., $k_{\textrm{B}} T < \hbar \omega_{\perp}$ with the Boltzmann constant $k_{\textrm{B}}$ and the reduced Planck constant $\hbar$.

The dynamics of the resulting cigar-shaped quasi-1D condensate of $N$ particles can be effectively described by a non-polynomial Schrödinger equation (npSE) (see, e.g., \cite{salasnich_effective_2002}) for its longitudinal wave function $\Psi(z,t)$ in normalized form
\begin{align}\label{eq:NPSE}
	i \partial_{t} \Psi(z,t) = &- \frac{1}{2m} \partial_{zz} \Psi(z,t) + V(z,t)\Psi(z,t) \\
	&+ \omega_{\perp} \left(\frac{1+3a_{s} N |\Psi(z,t)|^2}{\sqrt{1+2a_{s} N |\Psi(z,t)|^2}} - 1\right)\Psi(z,t),\nonumber
\end{align}
with the (normalized) mass $m$, the transversal scattering length $a_s$ and the transversal angular frequency $\omega_{\perp}$ given by the (normalized) transversal potential $V_{\perp}(x,y)$. The probability density $\rho(z,t)$ of finding a particle of the atomic cloud at position $z$ and at time $t$ can be obtained by $\rho(z,t) = |\Psi(z,t)|^2$ and thus
\begin{align}
	\int_{\mathcal{D}} |\Psi(z,t)|^2 \dd z = 1.
\end{align}
It is further assumed that $V(z,t)$ is sufficiently large at the domain boundaries such that $\Psi(\pm L/2,t)$ and $\partial_{z} \Psi(\pm L/2,t)$ vanishes.

\section{Optimal control of the non-polynomial Schrödinger equation}
\label{sec:optimal_control}

We assume that the spatial shape of the longitudinal potential $V(z,t)$ can be parametrized by some control parameter $\lambda(t) \in \mathcal{V}$, which is element of a suitable function space $\mathcal{V}$. 
The longitudinal potential is thus given by $V(z,t) = V_\lambda (z,\lambda(t))$, where the subscript highlights the dependence on $\lambda$ and thus implicitly on the time $t$. 

The aim of this section is to find trajectories $\lambda(t)$ that transfer the npSE from an initial state $\Psi_0(z) = \Psi(z,0)$, which is assumed to be the ground state of the initial potential $V(z,0)$, to a desired state $\Psi_{\text{des}}(z)$ at time $T$. While this can be an arbitrary state in general, we restrict ourselves to ground states of the final potential $V(z,T)$. We thus seek for what is commonly referred to as a "short-cut to adiabaticity", e.g., \cite{campo_shortcuts_2012}.

It is in general unclear if a trajectory for $\lambda$ exists such that the final state reaches $\Psi(z,T) = \Psi_{\text{des}}(z)$ exactly or whether this control problem is solvable within a given time horizon $T$. Therefore, the existing literature (e.g., \cite{winckel_computational_2008,grond_optimal_2009}) applies optimal control theory, where the optimality of such a transition is evaluated by a cost functional $J : \mathcal{V} \rightarrow \mathbb{R}$ in Bolza form
\begin{equation}
	J(\lambda) = \phi(T,\Psi(z,T)) + \int_{0}^{T}l(t,\Psi(z,t),\lambda(t))\dd t,
\end{equation}
with the terminal costs $\phi(T,\Psi(z,T))$ and the integral costs $l(t,\Psi(z,t),\lambda(t))$ during the transition. An optimal trajectory\footnote{Variables at the optimum are marked with a hat, which is not to be confused with operators in quantum mechanics.} $\hat{\lambda}(t)$ is thus given by the constrained optimization problem
\begin{subequations}\label{eq:optimization_problem}
	\begin{align}
		\min_{\lambda(\cdot)} &\,\,J(\lambda)\\%, \: \lambda :  (0, T) \rightarrow \mathbb{R},\\
		\text{s.t. }	i\partial_t  \Psi(z,t) &= f(\Psi(z,t),\lambda(t)), \label{eq:system}\\
		\Psi(z,0)  &= \Psi_0(z), \label{eq:psi_bound}\\
		\lambda(0) &= \lambda_0,\:\lambda(T) = \lambda_T, \label{eq:lambda_bound}
	\end{align}
\end{subequations}%
where $f(\Psi(z,t),\lambda(t))$ is used as an abbreviation of the right-hand side of \prettyref{eq:NPSE}.
\subsection{State and energy cost functionals}
A typical choice of the cost functional $J(\lambda)$ evaluates the difference between the obtained final state $\Psi(z,T)$ and the desired state $\Psi_\text{des}(z)$, i.e., 
\begin{align}
	J_\text{s} & = \frac{1}{2} \left( 1 - \left|  \int_\mathcal{D} \Psi^*_\text{des}(z) {\Psi}(z,T) \dd z \right|^2 \right) \label{eq:J_state}\\ 
	& \quad+  \frac{\gamma_\text{reg}}{2} \int_{0}^{T}  (\partial_t\lambda(t))^2 \dd t, \nonumber
\end{align}
with $\gamma_\text{reg} > 0$ and the complex conjugate $(\cdot)^*$. The term $\frac{\gamma_\text{reg}}{2} \int_{0}^{T}  (\partial_t\lambda(t))^2 \dd t$ penalizes strong variations of $\lambda(t)$ to regularize the optimization problem, improve the convergence (see \cite{winckel_computational_2008}) and ensure smooth trajectories of $\lambda$ which are favorable for experimental implementation.

If the desired state $\Psi_{\text{des}}(z)$ is the ground state of the final potential $V(z,T)$, one can alternatively consider the energy-based cost functional of the form
\begin{align}
	J_\text{e} = \int_\mathcal{D}  \mathcal{H}\left(\Psi(z,T)\right)\dd z - J_\text{e,des} + \frac{\gamma_\text{reg}}{2} \int_{0}^{T}  (\partial_t\lambda(t))^2\dd t, \label{eq:J_energy_npSE}
\end{align}
with $\gamma_\text{reg} > 0$ and
\begin{align}
	J_\text{e,des} &= \int_\mathcal{D} \mathcal{H}\left(\Psi_\text{des}(z,T)\right)\dd z
	\\ \intertext{using the Hamiltonian density, see \cite{erne_2018},}
	\mathcal{H}\left(\Psi(z,t)\right) &=  -\frac{1}{2m} \Psi^*(z,T) \partial_{zz} \Psi(z,t) + V_\lambda(z,\lambda_T)|\Psi(z,t)|^2\nonumber\\ %-\mu\Psi(z,t) \nonumber\\
	&\quad +\omega_\bot \sqrt{1+2a_sN |\Psi(z,t)|^2}  |\Psi(z,t)|^2.
\end{align}
Since the ground state is by definition the state of lowest energy, both the state and the energy cost functionals have the same global minimum in this case. Note that both cost function have the same value at this global minimum by construction in the absence of degenerate states. The convergence behavior of the two cost functionals, however, can still differ substantially, as we will see in \prettyref{sec:compression}.

\subsection{Optimality conditions} % better title, lagragian ...
To find the optimal trajectory $\hat{\lambda}(t)$ without restricting the search space apriori, one can apply methods from the calculus of variations to obtain a set of equations that are necessary conditions for optimality (sometimes called optimality system). 
Using the G\^ateaux derivative with respect to the control parameter
\begin{equation}
	\delta J(\lambda; \xi_\lambda) = \Diff{}{\nu} J(\lambda+\nu \xi_\lambda)\vert_{\nu = 0}
\end{equation}
where $\xi_\lambda \in \mathcal{V}$ is a variation of $\lambda$, a minimum of $J(\lambda)$ has to meet the first-order optimality condition
\begin{equation}\label{eq:first_order_optimality}
	\delta J(\hat{\lambda}; \xi_\lambda)=0
\end{equation}
for all admissible variations $\xi_\lambda$ observing the boundary conditions \eqref{eq:lambda_bound}. Note that the cost functional $J$ depends on the wave function $\Psi(z,t)$ and thereby on the control parameter through the system dynamics \prettyref{eq:NPSE}.

A commonly used approach is to follow the method of Lagrangian multipliers by introducing the Lagrange functional 
\begin{align}
	&L(\lambda,\Psi,p) = \phi(T,\Psi(z,T)) + \int_{0}^{T}l(t,\Psi(z,t),\lambda(t))\dd t \label{eq:L}\\ 
	&+ \real{\int_{0}^{T}\!\!\!\int_\mathcal{D}\! p^*(z,t)\left(\partial_t  \Psi(z,t) -\! f(\Psi(z,t),\lambda(t))\right) \dd z \dd t}, \nonumber
\end{align}
which uses the system dynamics \prettyref{eq:system} together with the adjoint variable $p(z,t)$.
With this definition, the condition \eqref{eq:first_order_optimality} is equivalent to vanishing (partial) G\^ateaux derivatives of $L(\lambda,\Psi,p)$ with respect to $\lambda$, $\Psi$, and $p$ (cp. \cite{winckel_computational_2008}), i.e.,
\begin{subequations}\label{eq:lagrange_optimality}
\begin{align}
	\delta_{p}L(\lambda,\Psi,p; \xi_p) &= 0 \label{eq:langrange_optimality_p}\\
	\delta_{\Psi} L(\lambda,\Psi,p; \xi_\Psi)&={0} \label{eq:langrange_optimality_psi}\\
	\delta_{\lambda} L(\lambda,\Psi,p; \xi_\lambda) &= 0 \label{eq:langrange_optimality_lambda}
\end{align}
\end{subequations}
for all admissible variations $\xi_\lambda$, $\xi_p$, and $\xi_\Psi$.% regarding the boundary conditions \eqref{eq:psi_bound} and \eqref{eq:lambda_bound}
%With this definition, $\delta J(\lambda; \xi_\lambda)$ is now equal to the partial G\^ateaux derivative  $\delta_{\lambda} L(\lambda,\Psi,p; \xi_\lambda)$ if the conditions  $\delta_{p}L(\lambda,\Psi,p; \xi_p)={0}$ and  \nolinebreak{$\delta_{\Psi} L(\lambda,\Psi,p; \xi_\Psi)={0}$} are met for all admissible variations of $\xi_p$ and $\xi_\Psi$, which are variations of the adjoint variable and the wave-function. This relation can be seen in the expansion
%\begin{equation}
%	\label{eq:var_J}
%	\delta J\left(\lambda; \xi_\lambda\right) = \delta_{\lambda} J\left(\lambda, \Psi; \xi_{\lambda}\right) + \delta_{\Psi} J\left(\lambda, \Psi; \xi_\Psi\right)|_{\xi_\Psi=\delta\Psi},
%\end{equation}
%where the derivative of the wave-function with respect to the control parameter $\delta\Psi(\lambda; \xi_\lambda)$ is derived by differentiating the state equation with respect to the control parameter
%\begin{equation*}
%	\partial_t \delta\Psi(\lambda; \xi_\lambda) = \frac{\partial}{\partial\Psi}f(\Psi,\lambda) \delta\Psi(\lambda; \xi_\lambda) + \frac{\partial}{\partial\lambda}f(\Psi,\lambda) \xi_\lambda
%\end{equation*}
%considering the initial condition 
%\begin{equation*}
%	\delta\Psi(\lambda; \xi_\lambda)|_{t=0} = {0}.
%\end{equation*}
%Setting the partial functional derivatives of \eqref{eq:L} with respect to $\lambda, \Psi$ and $p$ to zero results in a set of differential equations with initial and final conditions, which is also called optimality system and defines the first-order optimality condition.

Calculating \eqref{eq:lagrange_optimality} for all admissible variations regarding the boundary conditions \eqref{eq:psi_bound} and \eqref{eq:lambda_bound}, using the assumption that $V(z,t)$ is sufficiently large such that $\Psi(\pm L/2,t) = \partial_{z} \Psi(\pm L/2,t) = 0$, and applying the fundamental lemma of the variational calculus, we obtain the optimality system for the optimization problem \prettyref{eq:optimization_problem} as
%Thus, by setting  all partial G\^ateaux derivatives of $L(\lambda,\Psi,p)$ to zero for all admissible variations regarding the boundary conditions \eqref{eq:psi_bound} and \eqref{eq:lambda_bound}, using the assumption that $\rho(\pm L/2,t) = 0$ and applying the fundamental lemma of the variational calculus, we obtain the optimality system for the optimization problem \prettyref{eq:optimization_problem} as
\begin{subequations}
	\label{eq:opt_npSE}
	\begin{align}
		i  \partial_t\hat{\Psi}(z,t) =& \,f\big(\hat{\Psi}(z,t),\hat{\lambda}(t)\big) \label{eq:opt_npSE_stateEQ}\\
		%-\frac{1}{2m} \Delta \hat{\Psi}(z,t) + V_{\hat{\lambda}}(z,t)\hat{\Psi}(z,t) +\nonumber 
		%\\&+  \omega_\bot\left( \frac{1+3a_s N |\hat{\Psi}(z,t)|^2}{\sqrt{1+2a_sN|\hat{\Psi}(z,t)|^2}}+1\right)\hat{\Psi}(z,t)\label{eq:opt_Psi_npSE}\\
		i \partial_t  \hat{p}(z,t) =&  \left(-\frac{1}{2m} \Delta + V_{\lambda}(z,\hat{\lambda}(t)) +  A\big(\hat{\Psi}\big)\right)\hat{p}(z,t) \nonumber \\ 
		& +  B\big(\hat{\Psi}\big)\hat{p}^*(z,t) \label{eq:opt_npSE_adjointEQ}\\
		\gamma_\text{reg} \partial_{tt} \hat{\lambda}(t) =& - \real{\int \hat{\Psi}^*(z,t) \PartDiff{V_\lambda}{\lambda}}\bigg\vert_{\lambda=\hat{\lambda}(t)} p(z,t) \dd z \label{eq:opt_npSE_inputEQ}
	\end{align}
	whereby
%	\begin{align}
%		A(\Psi) =& \omega_\bot\Bigg(\frac{3a_sN|{\Psi}(z,t)|^2}{\sqrt{1+2a_sN|{\Psi}(z,t)|^2}} + \frac{1+3a_sN|{\Psi}(z,t)|^2}{\sqrt{1+2a_sN|{\Psi}(z,t)|^2}} \nonumber\\
%		& -\frac{(1+3a_sN|{\Psi}(z,t)|^2) a_s N |{\Psi(z,t)}|^2}{\left({1+2a_sN|{\Psi(z,t)}|^2}\right)^\frac{3}{2}} - 1\Bigg)\\
%		B(\Psi) =&  \omega_\bot\Bigg(\frac{3a_sN{\Psi}(z,t)^2}{\sqrt{1+2a_sN|{\Psi}(z,t)|^2}} \\ 
%		&\quad -\frac{(1+3a_sN|{\Psi(z,t)}|^2) a_sN {\Psi}(z,t)^2}{\left({1+2a_sN|{\Psi(z,t)}|^2}\right)^\frac{3}{2}}\Bigg). \nonumber
%	\end{align}
	\begin{align*}
	A(\Psi) =& \omega_\bot\Bigg(\frac{1+6a_sN|{\Psi}(z,t)|^2}{\sqrt{1+2a_sN|{\Psi}(z,t)|^2}} - 1 \\
	& -\frac{(1+3a_sN|{\Psi}(z,t)|^2) a_s N |{\Psi(z,t)}|^2}{\left({1+2a_sN|{\Psi(z,t)}|^2}\right)^\frac{3}{2}} \Bigg) \nonumber \\
	B(\Psi) =&  \omega_\bot\Bigg(\frac{3a_sN{\Psi}(z,t)^2}{\sqrt{1+2a_sN|{\Psi}(z,t)|^2}} \\ 
	&\quad -\frac{(1+3a_sN|{\Psi(z,t)}|^2) a_sN {\Psi}(z,t)^2}{\left({1+2a_sN|{\Psi(z,t)}|^2}\right)^\frac{3}{2}}\Bigg). \nonumber
	\end{align*}
	The boundary conditions for \prettyref{eq:opt_npSE_stateEQ} and \prettyref{eq:opt_npSE_inputEQ} are given by
	\begin{align}
		\hat{\Psi}(z,0) =& \Psi_0(z), \label{eq:opt_npSE_stateBC}\\
		\hat{\lambda}(0) =& \lambda_0, \qquad \hat{\lambda}(T) = \lambda_T. \label{eq:opt_npSE_inputBC}
	\end{align}
	The remaining terminal condition for the adjoint variable $p(z,t)$ at $t=T$ is given by either
	\begin{align}
		\hat{p}(z,T) =& i\Psi_\text{des}(z)\int_\mathcal{D} {\Psi}_\text{des}^*(z)\hat{{\Psi}}(z,T)\dd z \label{eq:opt_npSE_adjointBC_JS}\\
		\intertext{when using the state cost functional $J_\textrm{s}$ in \prettyref{eq:J_state} or by}
		\hat{p}(z,T) =& -2i\Bigg[ -\frac{1}{2m} \Delta  + V_{\lambda}(z,\lambda_T) \nonumber\\
		&+ \omega_\bot\left( \frac{1+3a_s N |\hat{\Psi}(z,T)|^2}{\sqrt{1+2a_sN|\hat{\Psi}(z,T)|^2}}+1 \right) \Bigg]\hat{\Psi}(z,T) \label{eq:opt_npSE_adjointBC_JE}
	\end{align}
	when using the energy cost functional $J_\textrm{e}$ in \prettyref{eq:J_energy_npSE}, respectively.
\end{subequations}
Note that solutions of \prettyref{eq:opt_npSE} only guarantee local optimality and suitable globalization strategies can be applied if needed.

\subsection{Solution of the optimality system} \label{sec:solution_optimality}
The respective optimality system can be solved using a gradient-based approach analogous to \cite{winckel_computational_2008}. 
Therefore, a gradient $\nabla J_\lambda$ is defined to meet
\begin{equation}
	\left( \nabla J_\lambda, \xi_\lambda\right)_X = \delta J(\lambda; \xi_\lambda)\label{eq:gradient}
\end{equation} 
for all admissible directions $\xi_\lambda \in X$ and the inner product $(.,.)_X$ of the inner-product space $X$. Typical choices include the Sobolev $H^1$ or the Lebesgue $L^2$ space using $({a},{b})_{L^2} = \int_0^{T}a(t)b(t) \dd t$ and $({a},{b})_{H^1} = \int_0^{T}\partial_ta(t)\partial_tb(t)\dd t$. %Note that the $H^1$ inner product lacks the term $\int_0^{T}a(t)b(t) \dd t$ from the usual definition and the Dirichlet boundary conditions of $\lambda(t)$ will impose boundary conditions for $\nabla J_\lambda$. 
Following \cite{winckel_computational_2008}, using $H^1$ yields less oscillating and more robust solutions with respect to the choice of $\gamma_\text{reg}$ while also attaining better cost-value results in less computation time. For this reason, $X = H^1$ chosen as inner product space in this work. 

Since $\delta J(\lambda; \xi_\lambda) = \delta_{\lambda} L(\lambda,\Psi,p; \xi_\lambda)$ if \eqref{eq:langrange_optimality_p} and \eqref{eq:langrange_optimality_psi} vanish, \eqref{eq:gradient}  results in a Poisson equation for $\nabla J_\lambda(t)$, i.e.
\begin{subequations}
	\label{eq:gradient_J}
	\begin{align}
		\DDiffT{}\nabla J_\lambda(t) &= \gamma_\text{reg} \ddot{{\lambda}}(t) + \real{\int_\mathcal{D} {\Psi}^*(z,t) \PartDiff{V_\lambda}{\lambda} p(z,t) \dd z},\\
		\intertext{with the boundary conditions}
		\nabla J_\lambda(0) &= 0, \qquad \nabla J_\lambda(T)=0.
	\end{align}
\end{subequations}

To obtain numerical solutions, the system dynamics \eqref{eq:opt_npSE_stateEQ} and the adjoint dynamics \eqref{eq:opt_npSE_adjointEQ} together with the terminal condition \prettyref{eq:opt_npSE_adjointBC_JS} or \prettyref{eq:opt_npSE_adjointBC_JE}, respectively, are solved by forward and backward integration starting from an initial guess of the control parameter $\lambda(t)$. This is done by employing a Crank-Nicolson scheme. Note that the implicit nature of the Crank-Nicolson scheme requires that a nonlinear system of equations is solved in each time step for the nonlinear state dynamics \eqref{eq:opt_npSE_stateEQ} unlike for the linear adjoint dynamics \eqref{eq:opt_npSE_adjointEQ}.
% Since \eqref{eq:adjoint_npSE} is linear in $p$, no underlying iterations are needed to solve the implicit equation of the Crank-Nicolson scheme.
With the values of  $\Psi$ and $p$ from the forward and backward integration, the two-point boundary value problem \prettyref{eq:gradient_J} can be solved directly by spatial discretization.
Note that $\nabla J_\lambda$ vanishes at time $t=0$ and $t=T$ and thus is an admissible variation of the boundary conditions \eqref{eq:lambda_bound}. This is a feature of the $H^1$ space and would not be the case in the $L^2$ space setting, see \cite{winckel_computational_2008}.
With the obtained gradient $\nabla J_\lambda(t)$, Quasi-Newton (BFGS) iterations are performed until $||\nabla J_\lambda||_X$ is below a certain tolerance or a maximum number of iterations is reached.
\section{Mean-field optimal compression in optical box potentials}
\label{sec:compression}
While optical potentials can be arbitrarily shaped in principle, the large number of experimental iterations required by existing heuristic approaches, see, e.g., \cite{tajik_designing_2019}, limits their usability for dynamic (i.e., time-varying) applications. However, these methods can be utilized to create optical box potentials by statically compensating for potential roughnesses of the bottom of the box while only dynamically moving the walls, which is easily possible. In the following, we want to illustrate the presented results for the compression of a quasi-1D Bose gas, which is a fundamental operation of a quantum field thermal machine.

\begin{figure}
	\centering
	\includegraphics[width=\columnwidth]{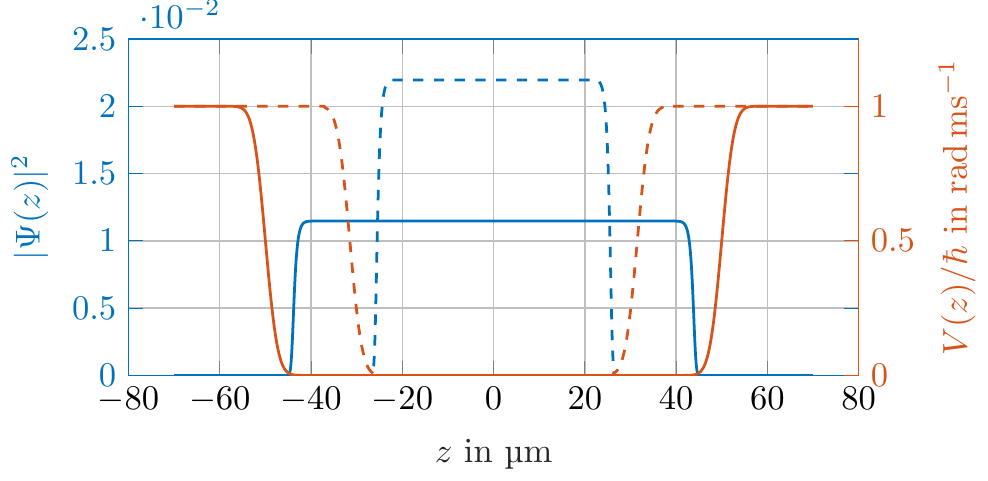}
	\vspace{-0.6cm}
	\caption{Initial and final potential and the density profiles of the corresponding ground states for the compression of a condensate from $\SI{100}{\micro m}$ to $\SI{50}{\micro m}$, i.e., $r_\text{comp} = 0.5$.}
	\label{fig:potential_groundstates}
\end{figure}

Since the steepness of optical box potentials is limited by the finite optical apperture, we assume that a box potential of initial length $w_0 < L$ is given by 
\begin{align}
	V_\lambda(z,\lambda(t)) &= V_\text{max} - \frac{V_\text{max}}{2}\erf\left(\frac{z+(w_0/2-\lambda(t))}{\sigma}\right) \nonumber\\
	&+ \frac{V_\text{max}}{2}\erf\left(\frac{z-(w_0/2-\lambda(t))}{\sigma}\right) \label{eq:Vt} 
\end{align}
using the Gaussian error function $\erf(z)$ with $\sigma = \SI{3}{\micro\metre}$, which is consistent with current experimental setups. Here, $\lambda(t)$ represents the displacement of the wall acting symmetrically on both sides as illustrated in \prettyref{fig:potential_groundstates}. For a final displacement $\lambda_T$, the resulting compression ratio is given by $r_\text{comp} = \frac{w_0-2\lambda_T}{w_0}$.

In the absence of an elaborate optimal control solution, experiments are typically performed using simple linear ramps for the control parameter, i.e., 
\begin{equation}
	\lambda_\textrm{lin}(t) = \left\{
	\begin{array}{ll}
		\lambda_0 + \frac{t}{T} (\lambda_T - \lambda_0) & t \in [0,T] \\
		\lambda_T & \, t \geq T,  \\
	\end{array}
	\right. \label{eq:lambda_lin}
\end{equation}
to achieve a dynamic transition of the applied potential.

% excitations travel with finite velocity - speed of sound, which increases with density
% contrary to scenarios where the entire potential landscape is actuated, moving potential walls solely rely on the condensate's dynamics to adjust a desired state. 
% similar to boundary control problem
% minimum control time c_s ~ 1.8 initially -> 75 / 1.8 = 42ms < t_min < 100 / 1.8 = 55ms
% T = 45ms
Dynamic changes of the control parameter $\lambda$ create excitations of the density that propagate with finite velocity $c_s$ usually called speed of sound. For the npSE, it is given by (see \cite{erne_2018})
\begin{equation}
	c_\text{s}^2 =  \omega_\bot a_s N \rho \frac{2+3a_s N \rho}{m(1+2a_s N \rho)^{3/2}}\label{eq:speed_of_sound_np}
\end{equation}
and assumes different values at each location $z$ in general due to its dependence on the local atomic density $\rho$. Contrary to scenarios where the entire potential landscape is actuated, a moving potential walls approach solely relies on the condensate's dynamics to act on neighboring particles and adjust a desired state. As a result, the time horizon $T$ for an optimal transition of the Bose gas cannot be shorter than the time it takes for excitations to reach the opposite wall of the box potential. The given scenario thus exhibits a minimum control time $T_\textrm{min}$ similar to boundary control problems, see, e.g., \cite{meurer_control_2013}. 
To obtain a choice for $T$ without exact results for the minimum control time, one can use the simple geometric estimate that 
\begin{equation}
	T_\textrm{min} \approx \frac{w_0}{c_\textrm{s,0}} \frac{ 1 + r_\textrm{comp}}{2},
\end{equation}
with the speed of sound $c_\textrm{s,0}$ corresponding to the initial flat density $\rho_0 = |\Psi_0|^2$ of the box potential. Note that this estimate ignores the fact that $c_s$ increases with density and thus typically overestimates the true minimum control time in a compression scenario.

\begin{figure}
	\centering
	\includegraphics[width=\columnwidth]{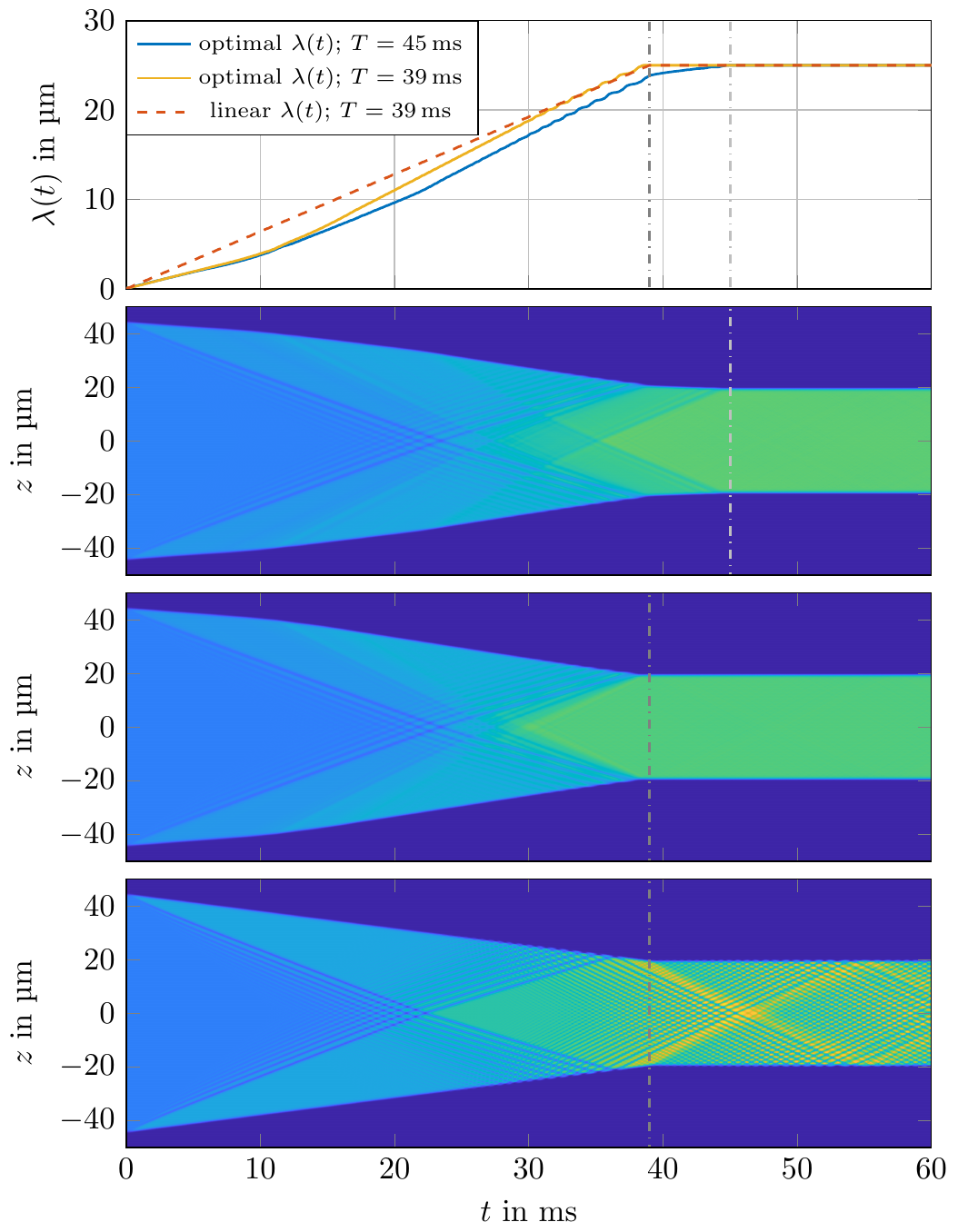}
	\vspace{-0.6cm}
	\caption{Comparison of a compression with $r_\text{comp} = 0.5$ using a linear ramp \prettyref{eq:lambda_lin} and optimal trajectories using the energy cost functional $J_\textrm{e}$ for $T = \SI{45}{\milli\second}$ and $T = \SI{39}{\milli\second}$. The top plot illustrates the evolution of the control parameter $\lambda(t)$. The corresponding evolution of the density $\rho(z,t)$ for the linear (optimal) trajectory of $\lambda(t)$ is shown in the plots below, whereby the optimal trajectories of $\hat{\lambda}(t)$ was determined by solving the optimality conditions \eqref{eq:opt_npSE} as described in \prettyref{sec:solution_optimality}.}
	\label{fig:carpet_oct}
\end{figure}

Using the parameter values from \prettyref{tab:parameters} and a compression ratio of $r_\textrm{comp} = 0.5$ yields $T_\textrm{min} \approx \SI{45.96}{\milli\second}$.
Solving the optimality system\eqref{eq:opt_npSE} for the energy-based cost function \eqref{eq:J_energy_npSE} with $\gamma_\textrm{reg} = \num{1e-5}$ over a time horizon $T = \SI{45}{\milli\second}$ yields the results shown in Fig.~\ref{fig:carpet_oct}. For the initial trajectory $\lambda(t)$ in the optimization scheme a linear ramp \eqref{eq:lambda_lin} was chosen.
After $t = T$, the control parameter $\lambda$ is held constant to observe the evolution of the system state after the transition.
The optimal transition is able to eliminate excitations when approaching $T$ and ultimately ends up in the desired gound state of the compressed potential. 
From visual inspection alone one clearly sees that a shorter time horizon is possible, which is verified by the optimal trajectory for $T= \SI{39}{\milli\second}$ shown in Fig.~\ref{fig:carpet_oct}.
The evolution of the density for a linear transition using \eqref{eq:lambda_lin} is finally given for comparison. It is interesting to note that optimal solutions close to the true $T_\textrm{min}$ yield particularly simple and smooth trajectories for $\hat{\lambda}(t)$.
%While the excitations of the condensate's dynamic are clearly present after the linear transition using \eqref{eq:lambda_lin}, the optimal transition is able to eliminate excitations when approaching $T$ and ultimately end up in the desired gound state of the compressed potential.

A popular alternative in quantum control applications to the indirect optimization approach (IOA) of using the optimality conditions \eqref{eq:opt_npSE} is to directly restrict the function space of the control parameter $\mathcal{V}$ using a finite number of basis functions. The optimal solution in this restricted space is then given by a simpler static optimization problem that does not use further information on the system (black-box optimization). Such a basis-function approach (BFA) can be quite powerful, e.g., by combining it with a randomisation of the basis functions, see \cite{doria_optimal_2011}.
For the desired transition between two steady states of the system, a particularly simple finite-dimensional parametrization of the control parameter is given by
\begin{align}
	\lambda(t) =
	\begin{cases}
		\lambda_\text{lin}(t) +\sum_{k=1}^{M} a_k \sin\left(k \pi \frac{t}{T_t}\right) \quad & t\in [0,T]\\
		\lambda_\text{lin}(t) \quad & \textrm{else}.
	\end{cases}\label{eq:lambda_BFA}
\end{align}
Due to the choice of harmonic basis functions on top of the linear ramp, the resulting static optimization problem is unconstrained apart from the system equation, i.e., the resulting $\lambda(t)$ is a feasible (smooth) trajectory for all possible coefficients $a_k$, $k = 1\ldots M$. To find the optimal coefficients $a_k$, one can use optimization algorithms that do not require (explicit) gradient information, which is essential for more involved quantum systems that are either impossible or difficult and time consuming to simulate. Together with the typically highly non-convex nature of the resulting static optimization problem, global optimization tools such as Bayesian or surrogate optimization methods are beneficial, see, e.g., \cite{jones_efficient_1998,frazier_tutorial_2018}. Due to the comparatively fast simulation time of the npSE, a Quasi-Newton algorithm without gradient information is used.
\begin{figure}
	\centering
	\includegraphics[width=\columnwidth]{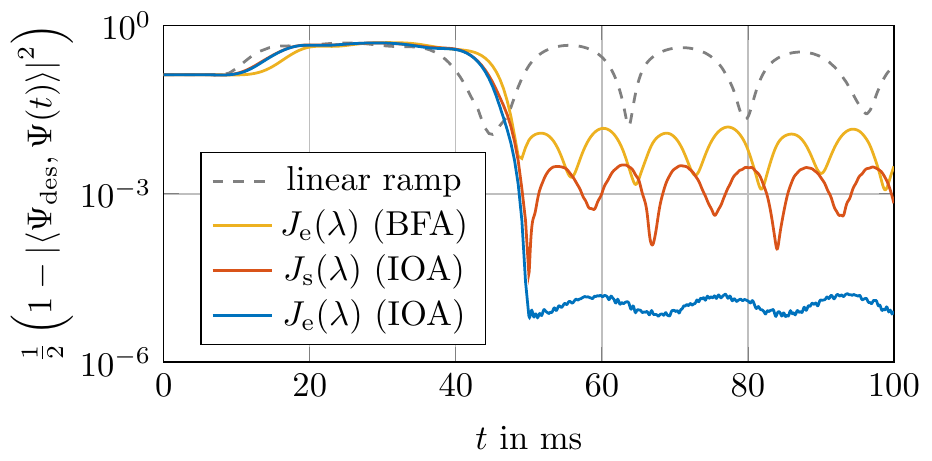}
	\vspace{-0.6cm}
	\caption{Evolution of the $L^2$-error $\frac{1}{2}\left( 1- |\langle \Psi_{\text{des}},\Psi(t) \rangle |^2 \right)$ of the system state $\Psi(t)$ of the resulting optimal $\hat{\lambda}(t)$ for different cost functionals $J_\textrm{e}(\lambda)$ and $J_\textrm{s}(\lambda)$ using the indirect optimization approach (IOA). The corresponding evolution when using a basis function approach (BFA) with $M = 4$ and a simple linear ramp $\lambda_\textrm{lin}(t)$ is given for comparison.}
	\label{fig:error_evolution}
\end{figure}

\begin{figure}
	\centering
	\includegraphics[width=\columnwidth]{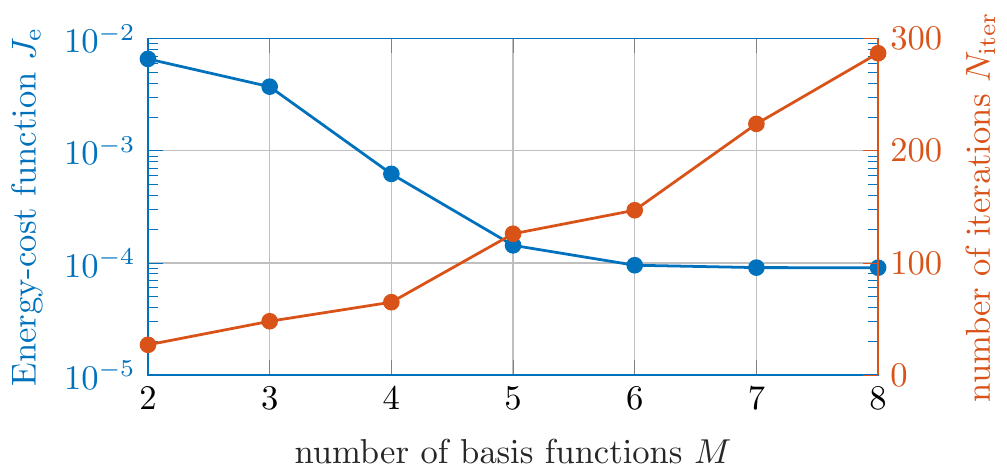}
	\vspace{-0.6cm}
	\caption{Resulting optimal values of the energy cost function $J_\textrm{e}(\hat{\lambda})$ for different number of basis functions $M$ and the required number of iterations $N_\textrm{iter}$ of the numerical optimization algorithm.}
	\label{fig:comp_number}
\end{figure}

To assess the resulting optimal control solutions for both cost functionals, i.e., state and energy costs, and indirect and basis function approaches, the evolution of the state error $\frac{1}{2} \left( 1 - \left|  \int_\mathcal{D} \Psi^*_\text{des}(z) {\Psi}(z,T) \dd z \right|^2 \right)$ is illustrated in \prettyref{fig:error_evolution} for $r_\textrm{comp} = 0.25$ and $T=\SI{45}{\milli\second}$. As one can see, the optimal solution using the energy cost function $J_\textrm{e}$ yields much better results. The reason for this is the significantly improved convergence behavior of $J_\textrm{e}$ compared to using $J_\textrm{s}$, which holds true for the IOA as well as the BFA. A  comparison of the obtained cost function values is given in \prettyref{tab:cost_75mum}. Not only are the required number of iterations $N_\textrm{iter}$ until the algorithms converges significantly lower, but also the state costs $J\textrm{s}$ of solutions obtained using  $J\textrm{e}$ are comparable or even lower than when using  $J\textrm{s}$ itself during the optimization.
The reduced search space somewhat limits the optimality of the resulting solutions and the achievable values of the cost function stagnate with an increasing number of basis functions, see \prettyref{fig:comp_number}, while the required number of iterations rises quickly.
\begin{table}
	\centering
	\begin{tabular}{p{2.2cm} c c >{\centering\arraybackslash}p{1.2cm}}
		& $J_\textrm{e}(\hat{\lambda})$ & $J_\textrm{s}(\hat{\lambda})$ & $N_\textrm{iter}$ \\
		\hline
		linear ramp & \num{2.9441e-2} & \num{1.9162e-1}&1\\
		BFA using $J_\textrm{e}$ & \num{6.2236e-4}&\num{8.9778e-3}&65\\
		BFA using $J_\textrm{s}$ & \num{2.7207e-2} & \num{2.4546e-3}&155\\
		IOA using $J_\textrm{e}$ & \num{1.5338e-5} & \num{6.7511e-6}&277\\
		IOA using $J_\textrm{s}$ & \num{5.3696e-4} & \num{3.7510e-5}&535\\
		\hline\\
	\end{tabular}
	\caption{Comparison of optimal trajectories $\hat{\lambda}$ found using either the IOA or BFA approach evaluated at both cost functionals $J_\textrm{e}$ and  $J_\textrm{s}$. For the BFA, $M = 4$ was chosen.}
	\label{tab:cost_75mum}
\end{table}

\begin{table}
	\centering
	\begin{tabular}{cl}
		Parameter & \text{Value}\\
		\hline
		normalized mass $m$ & \num{1.368} \\
		transversal frequency $\omega_\bot$  & \num{1e1} \\
		scattering length $a_s$ &\num{4.2e-3}\\
		particle number $N$ &\num{5.0e3} \\
		\hline
	\end{tabular}
	\vspace{0.1cm}
	\caption{Typical experimental parameters normalized to time in \si{\milli\second} and space in \si{\micro\meter}.}
	\label{tab:parameters}
\end{table}

%\begin{table}
%	\centering
% 	\begin{tabular}{clS}
%	Symbol & Name & \text{Value}\\
%	\hline
%	$m$ & normalized mass & 1.368 \\
%	$\omega_\bot$ & transversal confinement frequency &10.000 \\
%	$a_s$ & 	s-wave scattering length &4.200e-3\\
%	$N$ & atom number &5.000e3 \\
%	\hline
%	\end{tabular}
%	\vspace{0.1cm}
%	\caption{Typical normalized parameters from cold atom experiments.}
%	\label{tab:parameters}
%\end{table}

%\begin{table}
%	\centering
%	\begin{tabular}{clSc}
%		Parameter & \text{Value} & \text{Unit}\\
%		\hline
%		normalized mass $m$ & 1.3683 & \\
%		transversal frequency $\omega_\bot$  &10.000 & \si{\radian\per\milli\second}\\
%		scattering length $a_s$ &4.200e-3&\si{\micro \meter}\\
%		particle number $N$ &5.000e3 & 1\\
%		\hline
%	\end{tabular}
%	\caption{Typical parameters from cold atom experiments..}
%	\label{tab:parameters}
%\end{table}

\section{Conclusions and outlook}
\label{sec:outlook}

In this work, we presented an optimal control solution for the mean-field of a quasi-1D Bose gas that is described by a non-polynomial Schrödinger equation. Necessary optimality conditions for energy and state cost functionals were derived using a variational formulation and solutions of the resulting optimal system were used to optimally compress a quasi-condensate in an (optical) box potential. Energy cost functionals show superior convergence properties for the derived indirect optimization approach as well as for simple basis function approaches. The latter might be particularly useful beyond mean-field scenarios where accurate closed-form descriptions of the (stochastic) system dynamics are hardly available. Due to the particularly simple and smooth solutions for optimal choices of the time horizon, one could further include $T$ in the optimization for a tailored set of basis functions.

%===============================================================================
% Acknowledgements
\begin{ack}
This research was funded in part by the Austrian Science Fund (FWF) [P36236] and the DFG Research Unit FOR 2724 on “Thermal machines in the quantum world” (FWF I-6047). Financed by the European Union - NextGenerationEU. S.E. acknowledges an ESQ (Erwin Schrödinger Center for Quantum Science and Technology) fellowship funded through the European Union’s Horizon 2020 research and innovation programme under the Marie Sklodowska-Curie grant agreement No 801110.
\end{ack}

\bibliography{bibliography}             % bib file to produce the bibliography
                                                     % with bibtex (preferred)

% % Appendix ( if required )                                                   
% \appendix
% \section{A summary of Latin grammar}    % Each appendix must have a short title.
% \section{Some Latin vocabulary}              % Sections and subsections are supported  
%                                                                          % in the appendices.

\end{document}